\documentstyle[12pt,psfig]{article}

\def\d{^\dagger}
\def\e{{\rm e}}
\def\expect#1{\langle #1 \rangle}
\def\Tr{{\rm Tr}}
\def\bra#1{{\langle #1 |}}
\def\ket#1{{| #1 \rangle}}
\def\sx{{\hat \sigma}_x}
\def\sy{{\hat \sigma}_y}
\def\sz{{\hat \sigma}_z}
\def\sp{{\hat \sigma}_+}
\def\sm{{\hat \sigma}_-}
\def\H{{\hat H}}
\def\Heff{{\hat H}_{\rm eff}}
\def\T{{\hat T}}
\def\Teff{{\hat T}_{\rm eff}}
\def\Tenc{{\hat T}_E}
\def\Tdec{{\hat T}_D}
\def\L{{\hat L}}
\def\rhat{{\hat\rho}}
\def\one{\leavevmode\hbox{\normalsize1\kern-4.6pt\large1}}

\newcommand{\Encod}{{\sf E}}
\newcommand{\Decod}{{\sf D}}

\begin{document}

\title{Effects of noise on quantum error correction algorithms}

\author{Adriano Barenco\thanks{email:  barenco@physics.oxford.ac.uk} \\
\small \em  Clarendon Laboratory, Department of Physics, \\
\small \em  University of Oxford, Oxford  OX1 3PU, U.K. \smallskip \\
Todd A. Brun\thanks{email:  t.brun@qmw.ac.uk} \\
\small \em  Department of Physics, Queen Mary and Westfield College, \\
\small \em  University of London, London  E1 4NS, U.K. \\
\small \em  and \\
\small \em  Group of Applied Physics, University of Geneva, \\
\small \em  CH-1211 Geneva 4, Switzerland \smallskip \\
R\"udiger Schack\thanks{email:  r.schack@rhbnc.ac.uk} \\
\small \em  Department of Mathematics, Royal Holloway, \\
\small \em  University of London, \\
\small \em  Egham, Surrey TW20 0EX, U.K.  \smallskip\\
Timothy P. Spiller\thanks{email: ts@hplb.hpl.hp.com} \\
\small \em  Hewlett-Packard Laboratories, Bristol, \\
\small \em  Filton Road, Stoke Gifford, Bristol  BS12 6QZ, U.K.}

\maketitle
\newpage

\begin{abstract}
  It has recently been shown that there are efficient algorithms for
  quantum computers to solve certain problems, such as prime
  factorization, which are intractable to date on classical computers.
  The chances for practical implementation, however, are limited by
  decoherence, in which the effect of an external environment causes
  random errors in the quantum calculation.  To combat this problem,
  quantum error correction schemes have been proposed, in which a
  single quantum bit (qubit) is ``encoded'' as a state of some larger
  number of qubits, chosen to resist particular types of errors.  Most
  such schemes are vulnerable, however, to errors in the encoding and
  decoding itself.  We examine two such schemes, in which a single
  qubit is encoded in a state of $n$ qubits while subject to dephasing
  or to arbitrary isotropic noise.  Using both analytical and
  numerical calculations, we argue that error correction remains
  beneficial in the presence of weak noise, and that there is an
  optimal time between error correction steps, determined by the
  strength of the interaction with the environment and the parameters
  set by the encoding.
\end{abstract}

\section{Introduction}

Soon after the discovery of fast quantum algorithms for
factorization~\cite{Shor1}, it was realized that the efficiency of
quantum computers depends crucially on the control of errors during a
computation. This is not surprising in itself, since classical
computers also require an active monitoring of errors to operate
properly. However, the dissipative techniques used in classical error
correction destroy the superpositions necessary for quantum computation.
This problem stimulated an important effort in the direction of quantum
error correction. In the last year or so, after the initial discovery
of quantum error correction codes by Shor~\cite{shor2} and
Steane~\cite{st1,st2}, significant progress has been made in the
development and understanding of these codes. Much attention has been
devoted to the construction of codes using a variety of different
techniques to convert classical codes into quantum codes~\cite{st3,
  CS96} and providing a mathematical description of large families of
these codes~\cite{G96,CRS96}. Minimal codes, that
correct only one or a few errors, were also derived~\cite{laflamme,bennett}.
Most of this work addresses the issue of how most efficiently
to preserve a quantum state in a noisy environment {\em given\/} that
encoding and decoding can be done in an error-free way. Only a few
recent papers have addressed the possibility of encoding and decoding
in the presence of noise~\cite{sh3,divi}; these fault-tolerant schemes
are relatively complicated and involve many more qubits than the earlier
simple codes. It is therefore unlikely that we will see an
experimental implementation of these more elaborate proposals in the
near term. On the other hand, the issue of errors arising during
encoding and decoding has only been partially investigated in the
simplest error correcting codes proposed so far~\cite{chuang}. These
are the codes that could be implemented in a near-future quantum
computer.  The aim of this work is to bridge this gap, and provide
numerical as well as algebraic evidence that for certain regimes of
noise, error correction is worthwhile even when noise is present
during the encoding and decoding steps, as will be the case in any real
experiment. 

This paper is organized as follows.  Section~\ref{noiseandec} reviews
the effect of unwanted environmental coupling on a quantum computer,
its description in terms of master equations, and the fundamental
operation of error correcting codes. Section~\ref{analytic} presents
an analytical discussion of the effect of errors on an evolution
consisting of encoding, free evolution of an encoded qubit, and
decoding, with noise present at all stages.  The concept of a quantum
trajectory is described in section~\ref{sec4}, where numerical
simulation algorithms are presented using two unravelings of the
master equation (quantum jumps and quantum state diffusion) as well as
direct numerical solution of the master equation. The numerical
methods are found to be in reasonable agreement with the analytical
model.

\section{Noise and error correction}
\label{noiseandec}

\subsection{Decoherence}

A quantum system in complete isolation evolves according to
the Schr\"odinger equation
\begin{equation}
{{d\ket\psi}\over{dt}} = - {i\over\hbar} \H \ket\psi,
\end{equation}
where $\ket\psi$ is the state of the system (in this case a quantum
computer) and $\H$ is the Hamiltonian (in this case representing the
action of the quantum ``gates;'' in general, $\H$ will be
time-dependent).  This evolution is unitary.

Unfortunately, the approximation of a system being isolated is only
good for microscopic noninteracting systems.  As a system becomes
larger and more complicated, the effects of the environment become
more important.

Consider the example of a single qubit interacting with an external
environment.  The state of the qubit is described by a vector in a 
two-dimensional Hilbert space. A convenient basis is the canonical 
basis ${\cal B}=\{\ket0,\ket1\}$.
Suppose that the qubit is initially in a superposition
state $\alpha\ket0 + \beta\ket1$ and the environment in some unknown
state $\ket{A}$.  As the system and environment interact, the initial
product state $(\alpha\ket0 + \beta\ket1) \otimes \ket{A}$ can evolve
into an entangled state $\alpha\ket0 \otimes \ket{B_0} + \beta\ket1
\otimes \ket{B_1}$, where the environment has become correlated with
the state of the system (more realistic models of coupling with the
environment can be found in Ref~\cite{weiss}).  The system can no
longer be described by a state on its own.  Normally, an environment
is very complicated, containing many degrees of freedom, so it is
likely that $\ket{B_0}$ and $\ket{B_1}$ will be orthogonal (or very
nearly so). Thus, if we trace out the environment degrees of freedom,
an ensemble of our systems of interest is left in a {\it mixture},
described by a reduced density matrix
\begin{equation}
{\hat \rho} = |\alpha|^2 \ket0 \bra0
+ |\beta|^2 \ket1 \bra1.
\label{dephase}
\end{equation}
In effect, the environment has measured the value of the qubit, and
the superposition has been destroyed.  In this case, the evolution of
the reduced system is no longer unitary; and algorithms which depend
on the unitarity of the evolution, such as the Shor algorithm, will no
longer function.

The general effects of the environment on a quantum system can be very
complicated and difficult to describe.  However, a useful
approximation is to assume that the effects of the environment are
Markovian, or local in time.  In this case, it is possible to describe
the evolution of the reduced density matrix by a {\it master equation\/} of
Lindblad form \cite{Lindblad1976}
\begin{equation}
{d\over{dt}}\,\hat\rho =
 -{i\over\hbar}[\hat H,\hat\rho] +
 \sum_j\left(\hat L_j\hat\rho \hat L_j^\dagger
 - {1\over2} \hat L_j^\dagger\hat L_j\hat\rho   - {1\over2}\hat\rho
 \hat L_j^\dagger\hat L_j\right) \;,
\label{eqmaster}
\end{equation}
where $\hat H$ is the system Hamiltonian and the $\hat L_j$ are the
Lindblad operators representing the interaction with the environment.
This Markovian approximation is generally very good when the
environment is large compared with the system and the interaction
between them is fairly weak.  It might fail, however, for some realizations
of quantum computers.

What kinds of Lindblad operators typically occur in (\ref{eqmaster})?
This depends on the physics of the system, but certain operators are
common in quantum optical and atomic physics models.  One normal
effect of environmental interaction is {\it dissipation}, as in {\it
  spontaneous emission}.  If the qubit state $\ket1$ represents an
excited state, there will be a Lindblad operator proportional to the
lowering operator, of the form ${\hat L} = \sqrt{\kappa} {\hat
  \sigma}_-$, (${\hat \sigma}_- \ket1 = \ket0$ and ${\hat \sigma}_-
\ket0 = 0$). The qubit will tend to the ground state in the long
term, regardless of its initial state.  If the environment has a
non-zero temperature, there is the possibility of thermal excitations
as well, represented by another Lindblad operator proportional to the
raising operator ${\hat \sigma}_+ = {\hat \sigma}_{-}^{\dagger}$ .

Even if the rate of dissipation is small enough to be negligible, the
environment can still act to destroy superpositions.  As we saw in
(\ref{dephase}), correlations which develop with the environment can
randomly {\it dephase} the basis states $\ket0$ and $\ket1$. 
This process is represented by a Lindblad
operator proportional to the $z$ Pauli matrix, ${\hat L} = \sqrt{\kappa}
{\hat \sigma}_z$, (${\hat \sigma}_z \ket0 = - \ket0$ and ${\hat \sigma}_z
\ket1 = \ket1$).

The most general
interaction will reduce the qubit ensemble density operator to the one
at the center of the Bloch sphere. The individual (pure) states of the
members of the ensemble can be viewed as moving randomly on the
surface of the sphere. This effect is represented by {\it isotropic
  noise}, with three Lindblad operators proportional to ${\hat
  \sigma}_x, {\hat \sigma}_y$ and ${\hat \sigma}_z$, used in studying
the depolarizing channel \cite{G96,bennett}.  The exact choice of model
is determined by the physics of the quantum computer and its
environmental interactions.

\subsection{Error correcting codes}

For a qubit $\ket{\psi}$, the most
general form of single-qubit error induced by the environment can be written as
\begin{equation}
 \ket{\psi} \rightarrow {\hat E} \ket{\psi},
\label{error_form}
\end{equation}
where ${\hat E}$ is an arbitrary operator which can be decomposed as
\begin{equation}
  {\hat E}=e_1 \one + e_x {\hat \sigma}_x + e_y {\hat \sigma}_y + 
e_z {\hat \sigma}_z,
\label{generror}
\end{equation}
where $\one$ is the identity and the ${\hat \sigma}_i$ are the Pauli matrices.
In the simplest case (when one wants to protect a single qubit),
error correcting codes consist in encoding the basis states $\ket{0}$
and $\ket{1}$ of a qubit in well chosen states of several qubits:
\begin{equation}
\begin{array}{l}
 \ket{0} \rightarrow \ket{C_0} \\
 \ket{1} \rightarrow \ket{C_1},
\end{array}
\end{equation}
where $\ket{C_0}$ and $\ket{C_1}$ belong to the extended Hilbert space
of several qubits. Numerous techniques for constructing these codes
have appeared in the recent literature~\cite{quantph}.  These
error correcting techniques commonly assume that the encoding step
(i.e. the operation by which a single qubit in state $\alpha \ket{0} +
\beta\ket{1}$ is entangled with additional qubits to form the state
$\alpha \ket{C_0} + \beta \ket{C_1}$) as well as the decoding and
correcting steps are done in a noiseless environment. The issue of
noisy encoding and decoding has been addressed little outside the
context of fault-tolerant techniques, which require many more qubits
\cite{sh3,divi}.  In this work, we will focus on earlier and more
compact codes, and analyze the issue of noisy encoding and decoding.
In the next two sections we review the two codes that we have
analyzed.

\subsection{Dephasing noise}
\label{dephnoise}

\begin{figure}

\setlength{\unitlength}{0.030in}

\begin{picture}(75,60)(0,0)

\put(0,55){\mbox{a)}}

\put(10,15){\line(1,0){5}}
\put(25,15){\line(1,0){14}}
\put(10,30){\line(1,0){5}}
\put(25,30){\line(1,0){4}}
\put(10,45){\line(1,0){19}}

\put(15,25){\framebox(10,10){$U_A$}}
\put(15,10){\framebox(10,10){$U_A$}}

\put(30,44){\line(0,-1){13}}
\put(40,44){\line(0,-1){28}}

\put(30,30){\circle*{2}}
\put(30,45){\circle*{2}}
\put(40,45){\circle*{2}}
\put(40,15){\circle*{2}}

\put(31,30){\line(1,0){14}}
\put(31,45){\line(1,0){8}}
\put(41,45){\line(1,0){19}}
\put(41,15){\line(1,0){4}}

\put(45,25){\framebox(10,10){$U_A$}}
\put(45,10){\framebox(10,10){$U_A$}}

\put(60,10){\framebox(10,10){$U_y$}}
\put(60,25){\framebox(10,10){$U_y$}}
\put(60,40){\framebox(10,10){$U_y$}}

\put(70,15){\line(1,0){5}}
\put(70,30){\line(1,0){5}}
\put(70,45){\line(1,0){5}}

\put(55,15){\line(1,0){5}}
\put(55,30){\line(1,0){5}}
\put(2,15){$\ket{0}$}
\put(2,30){$\ket{0}$}
\put(2,45){$\ket{\psi}$}

\end{picture}
\hfill
\begin{picture}(88,60)(0,0)

\put(0,55){\mbox{b)}}

\put(10,15){\line(1,0){5}}
\put(10,30){\line(1,0){5}}
\put(10,45){\line(1,0){5}}

\put(15,10){\framebox(10,10){$U^\dagger_y$}}
\put(15,25){\framebox(10,10){$U^\dagger_y$}}
\put(15,40){\framebox(10,10){$U^\dagger_y$}}
\put(25,15){\line(1,0){5}}
\put(25,30){\line(1,0){5}}
\put(25,45){\line(1,0){19}}

\put(30,25){\framebox(10,10){$U_A$}}
\put(30,10){\framebox(10,10){$U_A$}}

\put(40,15){\line(1,0){4}}
\put(40,30){\line(1,0){14}}
\put(46,15){\line(1,0){14}}
\put(56,30){\line(1,0){4}}
\put(46,45){\line(1,0){8}}

\put(56,45){\line(1,0){19}}

\put(55,44){\line(0,-1){13}}
\put(45,44){\line(0,-1){28}}

\put(55,30){\circle*{2}}
\put(55,45){\circle*{2}}
\put(45,45){\circle*{2}}
\put(45,15){\circle*{2}}

\put(60,25){\framebox(10,10){$U_A$}}
\put(60,10){\framebox(10,10){$U_A$}}

\put(70,15){\line(1,0){5}}
\put(70,30){\line(1,0){5}}

\put(78,43){\mbox{$\ket{\phi}$}}
\put(78,28){\mbox{$\cal M$}}
\put(78,13){\mbox{$\cal M$}}

\end{picture}

\caption[fo0]{\small
  Encoding and decoding networks from~{\protect\cite{Braunstein}}. The
  graphical conventions are similar to those in
  Ref.~{\protect\cite{gang9}}.  The gate $U_A$ represents the
  operation $\ket{0} \rightarrow \ket{0}+\ket{1}$ and $\ket{1}
\rightarrow \ket{0}-\ket{1}$ (cf. Eq.~\ref{EquaA} in the text). The
gate $U_y$ is a $\pi$ rotation of the qubit along the $y$ axis
(Eq.~\ref{Equay}). The two-bit gates denoted by a line and two black
dots are ``control-phase'' gates. They change by $\pi$ the phase of a
quantum state only when both qubits are in state $\ket{1}$
(cf. Eq.~\ref{Equacphase}).}
\label{3bitfig}
\end{figure}
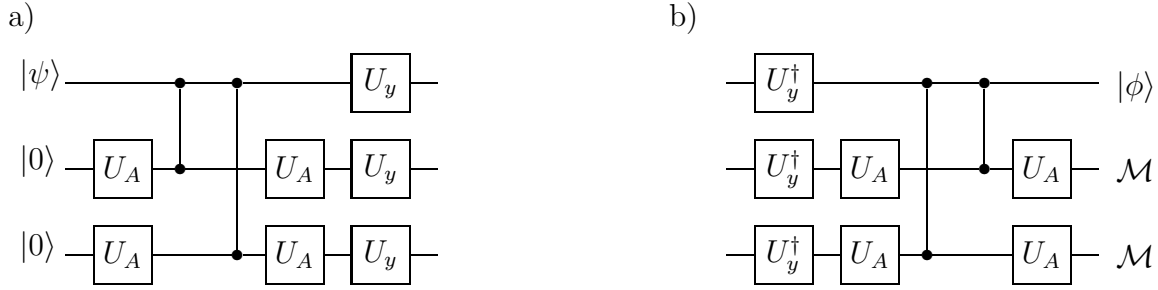

If one seeks to protect a qubit against a dephasing noise (in
Eq.~\ref{generror} this is equivalent to setting $e_y=e_z=0$), it can
be shown that the smallest possible code requires three qubits to
encode the states $\ket{0}$ and $\ket{1}$ of the initial qubit. This
carefully chosen superposition was first proposed by
Shor~\cite{shor2}. We use here an equivalent version found in
Ref.~\cite{st3,Braunstein}, in which
\begin{equation}
\begin{array}{l}
  \ket{0}\rightarrow \ket{C_0}=
  \ket{000}+\ket{001}+\ket{010}+\ket{011}+\ket{100}+\ket{101}+\ket{110}+\ket{111}
  \\ \ket{1}\rightarrow \ket{C_1}=
  \ket{000}-\ket{001}-\ket{010}+\ket{011}-\ket{100}+\ket{101}+\ket{110}-\ket{111}
\end{array}
\label{threebit}
\end{equation}
(normalization factors have been omitted).
We use the networks for encoding and for decoding/correcting shown in 
Fig.~\ref{3bitfig}. Initially, the first qubit is in the state
$\alpha\ket0 + \beta\ket1$.  This is the state we wish to protect.  The
second and third qubits are in the state $\ket{0}$.  The
result of the encoding network is
the three-qubit state $\alpha \ket{C_0} + \beta \ket{C_1}$.

In these figures we choose the various quantum gates in such way that
they can be described by simple Hamiltonians.  The gates $A$
correspond to the unitary operation
\begin{equation}
U_A=\frac{1}{\sqrt{2}}\left(
\begin{array}{cc}
1 &1\\
1&-1
\end{array}
\right)
\label{EquaA}
\end{equation}
effected by the Hamiltonian $H_A=\frac{\pi}{2}
(\frac{1}{\sqrt{2}} (\sx-\sz)+\one)$ acting for one unit
  of time. Similarly the gate $U_y$ corresponds to the unitary
  operation
\begin{equation}
U_y=\frac{1}{\sqrt{2}}\left(
\begin{array}{cc}
1 &1\\
-1&1
\end{array}
\right),
\label{Equay}
\end{equation}
generated by the Hamiltonian $H_y=-\frac{\pi}{4} \sy$
acting for one unit of time.

Please note that these matrices are represented in the basis
$\{ \ket0, \ket1 \}$, as is the convention in the quantum computation
literature.  Unfortunately, this is precisely the opposite of the usual
convention for the Pauli matrices.  For this paper we have retained the
quantum computation basis, as we do not present the Pauli matrices
explicitly, but this notational conflict should be resolved.

The two-bit gates of the network
correspond to controlled phase shifts.
These are represented in the canonical basis ${\cal B}=\{\ket{00},\ket{01},
\ket{10},\ket{11}\}$ by the unitary operator
\begin{equation}
U=\left(
\begin{array}{cccc}
1 & 0 &0 &0\\
0 & 1 &0 &0\\
0 & 0 &1 &0\\
0 & 0 &0 &-1\\
\end{array}
\right),
\label{Equacphase}
\end{equation}
generated by the Hamiltonian $H=\pi P_{\ket{1},i}\otimes
P_{\ket{1},j}$, where $i$ and $j$ designate the two qubits on which
the gate acts and $P_{\ket{1}}=\frac{1}{2}(\one + \sz)$ is the
projector on state $\ket{1}$. In this case a state $\ket{i,j}$ picks
up a phase $\pi$ iff both qubits are in state $\ket{1}$. Variants on
this gate can be obtained by replacing either or both of the
projection operators with $P_{\ket{0}}$ in the definition of the
Hamiltonian.

The decoding network is just the reverse of the encoding
network. After completing the sequence of gates, qubits 2 and 3 can
be measured to identify the error, which
is followed by an adequate correction of the first qubit~\cite{st3,
  Braunstein}.

\subsection{Arbitrary noise}

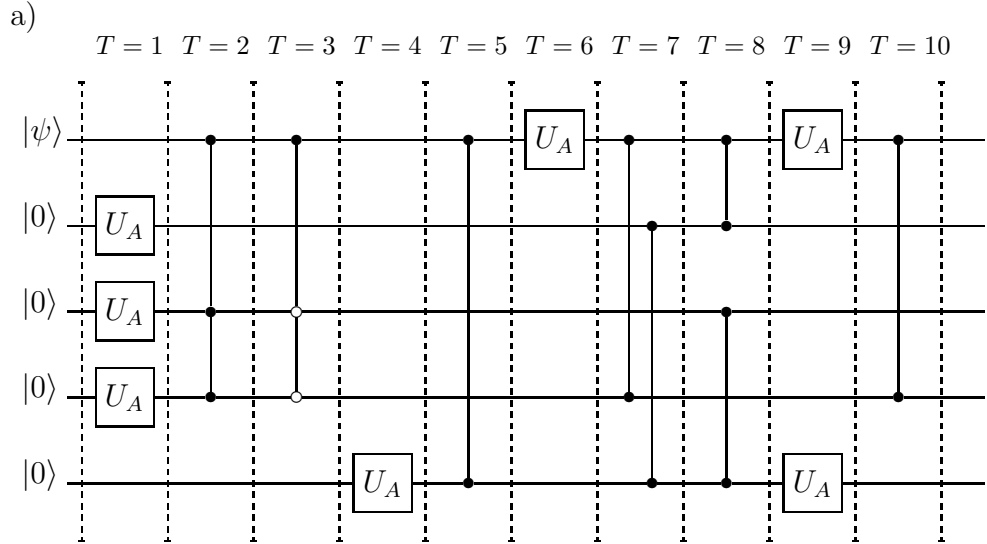
\begin{figure}

\setlength{\unitlength}{0.030in}

\begin{picture}(180,95)(0,0)

\put(0,95){\mbox{a)}}

\put(15,90){\mbox{\footnotesize$T=1$}}
\put(30,90){\mbox{\footnotesize$T=2$}}
\put(45,90){\mbox{\footnotesize$T=3$}}
\put(60,90){\mbox{\footnotesize$T=4$}}
\put(75,90){\mbox{\footnotesize$T=5$}}
\put(90,90){\mbox{\footnotesize$T=6$}}
\put(105,90){\mbox{\footnotesize$T=7$}}
\put(120,90){\mbox{\footnotesize$T=8$}}
\put(135,90){\mbox{\footnotesize$T=9$}}
\put(150,90){\mbox{\footnotesize$T=10$}}

\put(10,15){\line(1,0){26}}
\put(36,15){\line(1,0){15}}
\put(51,15){\line(1,0){9}}
\put(70,15){\line(1,0){9}}
\put(81,15){\line(1,0){30}}
\put(113,15){\line(1,0){11}}
\put(126,15){\line(1,0){9}}
\put(145,15){\line(1,0){11}}
\put(156,15){\line(1,0){14}}

\put(10,30){\line(1,0){5}}
\put(25,30){\line(1,0){9}}
\put(36,30){\line(1,0){13}}
\put(51,30){\line(1,0){56}}
\put(109,30){\line(1,0){45}}
\put(156,30){\line(1,0){14}}

\put(10,45){\line(1,0){5}}
\put(25,45){\line(1,0){9}}
\put(36,45){\line(1,0){13}}
\put(51,45){\line(1,0){73}}
\put(126,45){\line(1,0){44}}

\put(10,60){\line(1,0){5}}
\put(25,60){\line(1,0){86}}
\put(113,60){\line(1,0){11}}
\put(126,60){\line(1,0){44}}

\put(10,75){\line(1,0){24}}
\put(36,75){\line(1,0){13}}
\put(51,75){\line(1,0){28}}
\put(81,75){\line(1,0){9}}
\put(100,75){\line(1,0){7}}
\put(109,75){\line(1,0){15}}
\put(126,75){\line(1,0){9}}
\put(145,75){\line(1,0){9}}
\put(156,75){\line(1,0){14}}

\put(15,25){\framebox(10,10){$U_A$}}
\put(15,40){\framebox(10,10){$U_A$}}
\put(15,55){\framebox(10,10){$U_A$}}
\put(60,10){\framebox(10,10){$U_A$}}
\put(90,70){\framebox(10,10){$U_A$}}
\put(135,10){\framebox(10,10){$U_A$}}
\put(135,70){\framebox(10,10){$U_A$}}

\put(35,30){\circle*{2}}
\put(35,45){\circle*{2}}
\put(35,75){\circle*{2}}

\put(50,30){\circle{2}}
\put(50,45){\circle{2}}
\put(50,75){\circle*{2}}

\put(80,15){\circle*{2}}
\put(80,75){\circle*{2}}

\put(108,75){\circle*{2}}
\put(108,30){\circle*{2}}
\put(112,15){\circle*{2}}
\put(112,60){\circle*{2}}

\put(125,75){\circle*{2}}
\put(125,45){\circle*{2}}
\put(125,15){\circle*{2}}
\put(125,60){\circle*{2}}

\put(155,75){\circle*{2}}
\put(155,30){\circle*{2}}

\put(2,15){$\ket{0}$}
\put(2,30){$\ket{0}$}
\put(2,45){$\ket{0}$}
\put(2,60){$\ket{0}$}
\put(2,75){$\ket{\psi}$}

\put(35,74){\line(0,-1){28}}
\put(35,44){\line(0,-1){13}}

\put(50,74){\line(0,-1){28}}
\put(50,44){\line(0,-1){13}}

\put(80,74){\line(0,-1){58}}

\put(108,74){\line(0,-1){43}}
\put(112,59){\line(0,-1){43}}

\put(125,74){\line(0,-1){13}}
\put(125,44){\line(0,-1){28}}

\put(155,74){\line(0,-1){43}}

\put(12.5,5){\dashbox{1}(0,80){$$}}
\put(27.5,5){\dashbox{1}(0,80){$$}}
\put(42.5,5){\dashbox{1}(0,80){$$}}
\put(57.5,5){\dashbox{1}(0,80){$$}}
\put(72.5,5){\dashbox{1}(0,80){$$}}
\put(87.5,5){\dashbox{1}(0,80){$$}}
\put(102.5,5){\dashbox{1}(0,80){$$}}
\put(117.5,5){\dashbox{1}(0,80){$$}}
\put(132.5,5){\dashbox{1}(0,80){$$}}
\put(147.5,5){\dashbox{1}(0,80){$$}}
\put(162.5,5){\dashbox{1}(0,80){$$}}

\end{picture}

\caption[fo1]{\small
  Encoding network for the five bit error correcting code from
  Ref.~\cite{laflamme}. The network has been adapted in order to
  provide a more natural implementation of the gates in term of
  Hamiltonians.  The decoding network is identical to the encoding one
  read backwards. Dashed line indicate units of time, i.e., all the
  gates between two dashed line can be effected in one unit of time.
  The total encoding (or decoding) time is therefore 10 units. The
  gate that appear at time 2 is a generalization of the control-phase
  gate; it imparts a phase factor of $\pi$ to the quantum state only
  if all qubits denoted by a black dot are in state $\ket{1}$. It will
  effect the operation $\ket{1,i_2,1,1,i_5} \rightarrow
  -\ket{1,i_2,1,1,i_5}$ and leave all others states untouched. The
  gate at time 3 is a variation: the qubits indicated by a white dot
  act as controls when they are in state $\ket{0}$ rather than
  $\ket{1}$.}
\label{5bitfig}
\end{figure}

In the previous section we have shown how to protect a single qubit
against dephasing noise (i.e., noise generated by a single Lindblad
operator proportional to $\sz$). If one seeks to protect a single
qubit against an arbitrary error of the form (\ref{generror}), such as
isotropic noise, then five qubits are necessary to encode a state.
Different versions of these codes (all equivalent) have been
proposed~\cite{laflamme, bennett}.
We choose to implement an equivalent version of the 
code given by~\cite{laflamme}:
\begin{equation}
\begin{array}{ll}
  \ket{C_0}= &|b_1\rangle|00\rangle-|b_3\rangle|11\rangle+
              |b_5\rangle|01\rangle+|b_7\rangle|10\rangle\\
  \ket{C_1}=-&|b_2\rangle|11\rangle-|b_4\rangle|00\rangle-
              |b_6\rangle|10\rangle+|b_8\rangle|01\rangle, 
\end{array}
\label{5bitcode}
\end{equation}
where $|b_{1\atop 2}\rangle=(|000\rangle\pm|111\rangle)$, $|b_{3\atop
  4}\rangle=(|010\rangle\pm|101\rangle)$, $|b_{5\atop
  6}\rangle=(|001\rangle\pm|110\rangle)$, $|b_{7\atop
  8}\rangle=(|011\rangle\pm|100\rangle)$. The implementation of this
code is done in a way similar to the three bit case. 
In the first stage, a qubit in state $\ket{\psi_{in}}=\alpha \ket{0} +\beta
\ket{1}$ is entangled with four additional
qubits (initially in state $\ket{0}$) to produce the state $\alpha
\ket{C_0} + \beta \ket{C_1}$. This is done through the network of
Fig.~\ref{5bitfig}. Similar gates as in the previous section
are used. Note that some of the control-phase gates change the phase
when a qubit is in state $\ket{0}$ rather than in state $\ket{1}$,
unlike the three qubit code.

The decoding network is also given by Fig.~\ref{5bitfig} with the
gate operations performed in the reverse order (this is
possible because each gate appearing in the network is self-adjoint).
After decoding, qubits 2 to 5 are measured and, depending on the outcome, 
an appropriate correction is applied to return the first
qubit to the correct state.  (For a complete description of this code,
 see Ref.~\cite{laflamme}.)

\section{Analytic considerations}
\label{analytic}

Given that error correction can be implemented through encoding a one-qubit
state into an  $n$-qubit state, it is instructive to consider some
simple analytic conditions for the case of {\it imperfect} 
encoding, decoding and correction. These
conditions will complement our numerical results and indicate the parameter
ranges for which correction is potentially useful. 

Subsections \ref{ss31}-\ref{ss33} discuss a simple approach. 
To model the imperfect 
operation of the code, we assume that the decoherence which acts
during the encoding and decoding does not get corrected at all. Subsection
\ref{secerrorqj} examines the validity of modeling the influence of the
environment as instantaneous errors of the type (\ref{error_form}), and
demonstrates that this type of treatment can be made consistent with the
master equation (\ref{eqmaster}) to low order.

\subsection{Perfect single error correction}
\label{ss31}
First we need a benchmark. For simple decoherence, the 
probability that a single qubit {\it remains}
error free for a time $T$ is defined to be
\begin{equation}
p_{snc} = \e^{-\kappa_{n} T} \; .
\end{equation}
The ``$s$'' stands for success---it could be that the aim is to
successfully store the qubit for time $T$, or to transmit it down a
channel where the time taken for this is $T$---and ``$nc$'' indicates
that no correction procedure is applied to the qubit. If $\xi$
is the probability of an error and $\xi \ll 1$, then $\xi = 1 - p_{snc}
\approx \kappa_{n} T$.

The subscript $n$ on the decoherence rate $\kappa_{n}$ denotes the
fact that the number of qubits needed for encoding is determined by the type 
of noise. The relevant examples presented in the previous section
are $n=3$ for phase noise (modeled by $\hat L_{1} = \sqrt{\kappa} {\hat
  \sigma}_{z}$) and $n=5$ for isotropic noise (modeled by $\hat L_{1} =
\sqrt{\kappa} {\hat \sigma}_{x}$ , $\hat L_{2} = \sqrt{\kappa} {\hat
  \sigma}_{y}$ and $\hat L_{3} = \sqrt{\kappa} {\hat \sigma}_{z}$). We
define the mismatch between the ensemble at time $T$ and that $T=0$ to be
\begin{equation}
m_{nec}(T) \equiv 1 - \bra{\psi_{ini}} \rho \ket{\psi_{ini}},
\label{mismatchdef}
\end{equation}
where $\ket{\psi_{ini}}$ is the initial pure state at time $T=0$.  In
the phase noise example, an ensemble of single qubits given by
$\ket{\psi_{ini}} = 2^{-1/2} (\ket{0} + \ket{1})$ decoheres so that
the mismatch $m_{nec}$ is easily shown to be
\begin{equation}
  m_{nec}(T) = \frac{1}{2}\left(1 - \e^{-2 \kappa T}\right) \; .
\label{misphas}
\end{equation}
We therefore obtain $\kappa_{3} = 2 \kappa$, since the exponentially
decaying term in the mismatch can be identified with the probability
that the system remains error free. In the isotropic noise case, the
same initial ensemble decoheres and exhibits a mismatch of
\begin{equation} 
m_{nec}(T) = \frac{1}{2}\left(1 - \e^{-4 \kappa T}\right)
\label{misiso}
\end{equation}
and so we identify 
$\kappa_{5} = 4 \kappa$. The particular choice of initial state is made
so that it exhibits sensitivity to phase or to isotropic noise. It is 
also used in the numerical simulations, so we compare like with like.

Consider now the $n$-qubit encoding and decoding procedure which is
able to correct perfectly for a single error in one of the $n$ qubits,
but fails if there are two or more errors. Using this procedure, the
probability of the successful survival of a single encoded qubit state
for time $T$ is the sum of the zero error and one error probabilities;
\begin{equation}
p_{sc}(n) = \e^{-n \kappa_{n} T} \; + \; n \e^{-(n-1) \kappa_{n} T}\left(1 - 
\e^{-\kappa_{n} T}\right) \; .
\label{pperfect}
\end{equation}
(Each qubit is assumed to suffer the same decoherence rate $\kappa_{n}$.) 
Clearly, if $\kappa_{n} T \ll 1$, perfect 
error correction is worthwhile because $p_{sc}(n)$ is
closer to unity than is $p_{snc}$. ($p_{snc} \approx 1 - \kappa_{n} T$, 
whereas $p_{sc} \approx 1 - \frac{n(n-1)}{2} \kappa_{n}^{2} T^{2}$ .) 
For the case $n = 3$ the crossover point arises
when $p_{snc} = p_{sc}(3)$, which yields 
$\kappa_{3} T = \ln 2$. For the case $n = 5$, $p_{snc} = p_{sc}(5)$ gives 
$\kappa_{5} T = 0.14$. Any realistic systems are likely to be well down in
$\kappa_{n} T$ from these values and so would certainly benefit from perfect 
error correction.

\subsection{Imperfect error correction}
Consider now the case where the encoding (\Encod) and decoding
(\Decod) for the error correction procedure take a finite amount of
time. Let this be $T \delta$, so $\delta$ is the dimensionless
fraction of time taken by one full \Encod+\Decod\ stage. \Decod\ may be
essentially the inverse of \Encod\ and so each may take $\delta/2$,
but this is not crucial.  The decoherence rate could well differ
during \Encod\ and \Decod; we denote it by $\kappa_{n}^{\prime}$. The
environment seen by the qubits may be different when the encoding and
decoding interactions are occurring. The point to note is
that errors which occur during the \Encod+\Decod\ stage are unwelcome.
We assume that they don't get corrected and so contribute directly to
the error rate for qubit system.

If the problem at hand is the storage of a given qubit state for time
$T$, it seems reasonable to allow \Encod+\Decod\ to be part of $T$, so
the encoded $n$-qubit state is then kept for $(1 - \delta)T$.
Alternatively, if the goal is to propagate a qubit state down a
channel where the time for transmission is $T$, it would seem to be
more realistic to add on $T \delta$, so the whole process takes $(1 +
\delta)T$. This distinction does not appear to be crucial, but we
examine both cases.

\subsubsection{Storage with imperfect correction}
\label{storagesec}

The probability $s$ that there is no error in this system is the
product of the probability of all $n$ qubits surviving $T \delta$ at a
decoherence rate of $\kappa_{n}^{\prime}$ with the probability of zero or
one error (which can be corrected) during the time $(1 - \delta)T$ at
a rate $\kappa_{n}$. Using (\ref{pperfect}), this gives
\begin{equation}
s_{sc}(n) = \e^{-n \kappa_{n}^{\prime} T \delta} 
\left(n \e^{-(n-1) \kappa_{n} (1-\delta) T} 
\; - \; (n-1) \e^{-n \kappa_{n} (1-\delta) T}\right) \; .
\label{pstore}
\end{equation}
This reduces to $p_{sc}(n)$ given in (\ref{pperfect}) 
as $\delta \rightarrow 0$.

To make a simple comparison, let $\kappa_{n}^{\prime} = \kappa_{n}$. Equating
$s_{sc}(n)$ to $p_{snc}(n)$, the aim is to find the crossover value
for $\delta$.  Clearly $\delta$ is about $1/n$ ; the next order
correction in $\kappa_{n} T$ gives
\begin{equation}
  \delta = \frac{1}{n} \; - \; \frac{(n-1)^{3}}{2 n^{2}} \kappa_{n} T \; +
  \; ...
\end{equation}
Provided that $\delta$ stays below this value, there should be benefit
from error correction even though errors may occur during
\Encod+\Decod. However, if $\delta$ exceeds this value, the
performance of the procedure is actually worse than doing no
correction to a single qubit. For the cases $n = 3$ and $n = 5$ and
provided that $\kappa_{n} T \ll 1$, the bounds on $\delta$ are not very
constraining. Practical systems would probably have $\delta \ll 1$ and
so would operate in the regime where imperfect correction is
beneficial.

\subsubsection{Transmission with imperfect correction}
\label{transm}
The probability $t$ that there is no error in this system is the
product of the probability of all $n$ qubits surviving $T \delta$ at a
decoherence rate of $\kappa_{n}^{\prime}$ with the probability of zero or
one error (which can be corrected) during the transmission time $T$ at
a rate $\kappa_{n}$.  Using (\ref{pperfect}), this gives
\begin{equation}
  t_{sc}(n) = \e^{-n \kappa_{n}^{\prime} T \delta} \left(n \e^{-(n-1) 
\kappa_{n} T} \; - \; (n-1) \e^{-n \kappa_{n} T}\right) \; .
\label{ptrans}
\end{equation}
This also reduces to $p_{sc}(n)$ given in (\ref{pperfect}) as 
$\delta \rightarrow 0$.

To again make a simple comparison, let $\kappa_{n}^{\prime} = \kappa_{n}$. 
Equating 
$t_{sc}(n)$ to $p_{snc}(n)$, the aim is to find the crossover value for 
$\delta$. 
Once again $\delta$ is about $1/n$ ; the next order correction in 
$\kappa_{n} T$ gives
\begin{equation}
\delta = \frac{1}{n} \; - \; \frac{(n-1)}{2} \kappa_{n} T \; + \; ...
\end{equation}
The conclusion for transmission is the same as that for storage. Practical
systems with $\kappa_{n} T \ll 1$ and $\delta \ll 1$ will be in the regime 
where correction is beneficial.

\subsubsection{Single correction optimization}
\label{secres}

In the numerical simulations presented in Sect. \ref{sec4}, 
the encoding and decoding take a set amount of time,
rather than a set fraction of $T$. We therefore define an alternative
parameterization of the time taken for \Encod+\Decod, setting $T \delta
= \Delta$. Our analytic expressions can be viewed either in terms of
$\delta$ or of $\Delta$, whichever is most appropriate.

Here we just give a simple analytic result, optimizing the time $T$ to
achieve the most benefit from correction. Assuming that the time taken
for \Encod+\Decod\ is fixed at $\Delta$, what is the optimum $T$? We
find this by maximizing the ratio $R$
of the mismatch without correction to the 
mismatch with correction
\begin{equation}
  R \equiv \frac{1 - p_{snc}(n)}{1 - s_{sc}(n)} ,
\label{firstratio}
\end{equation}
and similarly for the transmission case. The maximum of $R$ is an indicator
of where (imperfect) error correction is giving the maximum benefit in
comparison to performing no correction at all; it is one of the
measures we use in our numerical work. To the lowest approximation
(assuming that $\kappa_{n} T \ll 1$ always), the optimum $T$ is the
same for both storage and transmission, and is given by
\begin{equation}
  T_{opt} \approx \left(\frac{2 \Delta}{(n - 1)
    \kappa_{n}}\right)^{1/2} \; .
\label{topt}
\end{equation}
Obviously, if the error correction is perfect (effectively taking zero
time), we arrive at the conclusion that it should be performed as
often as possible; $T_{opt} = 0$. However, for cases of practical
interest (finite $\Delta$) this is not so as $T_{opt}$ is then finite.
The dependence of $T_{opt}$ on $\kappa$ will be compared to our numerical
simulations in Fig.~\ref{logplot}.

\subsection{$N$-correction procedure}
\label{ss33}

The basic aim of error correction (within the context of this paper)
is to maximize the probability of success, storing or transmitting the
state as well as possible. The parameters $T$, $\kappa$,
$\kappa^{\prime}$, $n$ and $\delta$ are therefore set by the
problem at hand. $T$ is set by the total length of the transmission
channel or the total required storage time. (The latter might be the
time for which the state ``idles'' between interactions in a larger
quantum computation; for such a case the simple error correction
procedures discussed in this paper would be sufficient to keep it
coherent while it idles.) The decoherence is set by the environment.
$n$ and $\delta$ will be set by the chosen correction scheme and its
physical realization. However,  there is still some
freedom.  Given all the parameters above, the {\em number} of
correction procedures applied during $T$ can be varied.

\subsubsection{Perfect error correction}

Consider then the problem of optimizing error correction to achieve
the greatest probability of successful storage or transmission of a
qubit state, given the freedom to apply an arbitrary number $N$ of
\Encod+\Decod\ procedures during the time $T$.  Assume that these are
spaced out equally. For the case of perfect error correction, where
$\delta = 0$ so there is no possibility of an error occurring during
\Encod+\Decod, it is obviously beneficial to apply as many corrections
as possible. The probability of success for $N$ applications is
\begin{equation}
p_{Nsc}(n) = \left(n \e^{-(n-1) \kappa_{n} T/N} \; - \; 
(n-1) \e^{-n \kappa_{n} T/N}\right)^{N} \; .
\label{pNperfect}
\end{equation}
This maximizes for $N \rightarrow \infty$, tending to unity
independent of the value of $n$, and is consistent with our
observation that $T_{opt} = 0$ when $\Delta = 0$.  Such behavior is
like the ``Zeno'' or ``watchdog'' effect; there is no change at all
from the initial state as $N \rightarrow \infty$.

\subsubsection{Storage with imperfect correction}

In any realistic situation we will have $\delta > 0$. 
For any finite value of $\delta$, there is a non-vanishing
probability of introducing a non-correctable error for each
application of \Encod+\Decod, so it seems intuitively reasonable that
there should be an optimum value of $N$. Also, for the case of
storage, it makes sense to impose $N \delta < 1$, or else the time
taken for $N$ applications of \Encod+\Decod\ will exceed the time for
which the qubit is stored.  Practically, it
is likely that $N \delta \ll 1$ would hold.

The generalization of (\ref{pstore}) to $N$ equally spaced corrections is
\begin{equation}
  s_{Nsc}(n) = \e^{-n N \kappa_{n}^{\prime} T \delta} \left(n
  \e^{-(n-1) \kappa_{n} ((1/N)-\delta) T} \; - \; (n-1) \e^{-n
    \kappa_{n} ((1/N)-\delta) T}\right)^{N} \; .
\label{pNstore}
\end{equation}
For the simplest case of a decoherence rate always equal to $\kappa_{n}$,
equating the derivative (with respect to $N$) to zero, keeping only
the leading terms and rearranging to give the optimum $N$ yields
\begin{equation}
  N \approx \left(\frac{(n-1) \kappa_{n} T}{2 \delta}\right)^{1/2} =
  \left(\frac{(n-1) \kappa_{n}}{2 \Delta}\right)^{1/2} T \; .
\label{optN}
\end{equation}
Note that this is consistent with our result (\ref{topt}), if we
identify $T_{opt}$ with $T/N$.

\subsubsection{Transmission with imperfect correction}

Since the time for $N$ applications of \Encod+\Decod\ does not eat
into $T$, but adds to it for transmission, there is not the absolute
requirement that $\delta < (1/N)$. However, for practical cases it is
likely that $N \delta \ll 1$, the same as for storage. The
generalization of (\ref{ptrans}) to $N$ equally spaced corrections is
\begin{equation}
  t_{Nsc}(n) = \e^{-n N \kappa_{n}^{\prime} T \delta} \left(n \e^{-(n-1)
    \kappa_{n} T/N} \; - \; (n-1) \e^{-n \kappa_{n} T/N}\right)^{N} \; .
\label{pNtrans}
\end{equation}
The optimum $N$ is again given by (\ref{optN}), although there are
differences at the next order.

\subsubsection{$N$-step optimization}
As expected, there is an optimum number $N$ of corrections to apply
when these procedures themselves are imperfect. When $N \delta \ll 1$,
the optimum is the same for the storage and the transmission
scenarios.

It is interesting to substitute back the optimum $N$ of (\ref{optN}),
to obtain the maximum achievable transmission and storage success
probabilities. At the first order of approximation they are equal and
given by
\begin{eqnarray}
  \max_{N}(s_{Nsc}(n)) \approx \max_{N}(t_{Nsc}(n)) \approx
  \exp\left[-n \kappa_{n} T \left(2 (n-1) \delta \kappa_{n}
  T\right)^{1/2}\right] \nonumber \\ \approx 1 - n \kappa_{n} T \left( 2
  (n-1) \delta \kappa_{n} T\right)^{1/2} \; .
\label{maxp}
\end{eqnarray}
Thus, within the simple framework used here, the {\it minimum}
probability for a qubit state to incur an error in a total (storage or
transmission) time $T$ is approximately $n \kappa_{n} T \left( 2 \delta
(n-1) \kappa_{n} T\right)^{1/2}$. For cases of practical interest,
where this probability is small (and so the success probabilities are
close to unity), this is a good  approximation.

It is worth noting that, because of the square root in (\ref{optN}),
the optimum $N$ does not grow too quickly. For example, with
phase noise at $\kappa = 10^{-5}$ (i.e. $n = 3$ and $\kappa_{3} = 2
\kappa$), and with $\Delta = 10$ and a total time of $T =
10^{4}$, the minimum qubit error probability (calculated using
(\ref{maxp})) of 0.017 arises from applying $N \sim 14$ corrections.
With isotropic noise at the same $\kappa$ (i.e. $n = 5$ and $\kappa_{5}
= 4 \kappa$), and with $\Delta = 20$ and an elapsed time of $T =
10^{3}$, the minimum error probability of 0.016 is obtained with just
two corrections.

\subsection{Errors and quantum jumps}
\label{secerrorqj}

In all of this analysis we have been explicitly assuming that the
influence of the environment produces errors of type (\ref{error_form}).
It might be asked what the relation is between the general form of
single-qubit errors given in (\ref{error_form}) and the Lindblad
master equation (\ref{eqmaster}).  At first glance they seem to have
no resemblance to each other.  The former is an abrupt, instantaneous
change of state which occurs at random times; the latter is a
continuous, deterministic equation for the density operator $\rhat$.
This is the more correct description of the system's evolution.  In
what circumstances can we approximate it by (\ref{error_form})?

If we represent the right-hand side of equation (\ref{eqmaster})
by a superoperator
${\cal L}$, then the master equation becomes
\begin{equation}
\frac{d}{dt}{\rhat} = {\cal L}\rhat,
\end{equation}
and given the density operator $\rhat(0)$ at some initial time we can
formally solve for it a time $T$ later:
\begin{equation}
\rhat(T) = \e^{{\cal L} T}\rhat(0).
\end{equation}
We can expand the right-hand side of this equation to get
\begin{eqnarray}
\rhat(T) = && \e^{-i \Heff T} \rhat(0) \e^{i \Heff\d T} \nonumber\\
&& + \sum_j \int_0^T dt 
  \left[ \e^{-i \Heff (T-t)} \L_j \e^{-i \Heff t} \rhat(0) 
  \e^{i \Heff\d t} \L_j\d \e^{i \Heff\d (T-t)}\right] \nonumber\\
&& + \sum_j \sum_k \int_0^T dt \int_0^t dt'
  \left[ \e^{-i \Heff (T-t)} \L_j \e^{-i \Heff (t-t')}
  \L_k \e^{-i \Heff t'} \rhat(0) \right. \nonumber\\
&& \left. \times \e^{i \Heff\d t'} \L_k\d \e^{i \Heff\d (t-t')} \L_j\d
  \e^{i \Heff\d (T-t)} \right]
  + \cdots,
\label{rho_expansion}
\end{eqnarray}
where
\begin{equation}
  \Heff = \H - {i\over2}\sum_j \L_j\d \L_j
\end{equation}
is a non-Hermitian ``effective Hamiltonian''\cite{Carmichael1993b, 
Dalibard1992, Gardiner1992}. 

Already in this expansion we can see the relationship between the
stochastic model of errors (\ref{error_form}) and the continuous
master equation (\ref{eqmaster}) Each term in (\ref{rho_expansion})
looks like a collection of instantaneous ``jumps'' (or errors)
interrupting a continuous state vector evolution.

However, it should be noted that this continuous evolution is not
necessarily the {\it desired} evolution; the non-Hermitian component
may produce unwanted effects, depending on the Lindblad operators.  If
we choose dephasing noise, so that $\L = \sqrt{\kappa}\sz$, then the
effective Hamiltonian is $\Heff = \H - i (\kappa/2) \one$, resulting
merely in a renormalization of the state; in the case of spontaneous
emission, however, we have $\L = \sqrt{\kappa}\sm$ and $\Heff = \H - i
(\kappa/2) \sp\sm$, which changes the relative weight of the $\ket0$
and $\ket1$ states.

This sort of {\it continuous} error does not fit the error correction
paradigm, and therefore cannot be completely corrected.  However, all
is not lost: if $\kappa T$ is very small, then it is possible to
expand
\begin{equation}
  \e^{-i \Heff T} \approx \e^{-i \H T}(1 + \kappa T {\hat O}),
\end{equation}
where ${\hat O}$ is a function of the commutator of $\H$ and $\L$, and
to first order in $\kappa T$ the error correcting algorithm will still
work.

Similarly, we see that the second and higher terms in
(\ref{rho_expansion}) correspond to more than one ``error'' occurring
during the time $T$; hence, error correction techniques for single
errors will be ineffective for these terms.  But again, for small
$\kappa T$, these terms will be of higher order, so correction is
still beneficial.

The situation is somewhat more involved than this, however.  For most
models of a quantum computer the Hamiltonian will be time-varying.
Thus, the time-evolution operators will not be the simple exponentials
written in (\ref{rho_expansion}), but some more complicated operators.
In the simple case where a gate is effected merely by ``turning on''
some Hamiltonian for a set period of time and then ``turning it off''
again, the time evolution operator for the operation of $n$ gates
would be
\begin{equation}
\T(T,0) = \e^{-i\H_n t_n} \cdots \e^{-i \H_2 t_2} \e^{-i \H_1 t_1}.
\end{equation}
The expansion would then become
\begin{eqnarray}
\rhat(T) = && \Teff(T,0) \rhat(0) \Teff\d(T,0) \nonumber\\
&& + \sum_j \int_0^T dt 
  \left[  \Teff(T,t) \L_j \Teff(t,0) \rhat(0)
  \Teff\d(t,0) \L_j\d \Teff\d(T,t) \right] \nonumber\\
&& + \sum_j \sum_k \int_0^T dt \int_0^t dt'
  \left[\Teff(T,t) \L_j \Teff(t,t') \L_k \Teff(t',0)\rhat(0) \right. \nonumber\\
&& \left. \times \Teff\d(t',0) \L_k\d \Teff\d(t,t') \L_j\d
  \Teff\d(T,t) \right]
  + \cdots,
\label{rho_expansion2}
\end{eqnarray}
where the effective time-evolution operator $\Teff$ includes the
non-unitary effects of the environment, just as in
(\ref{rho_expansion}).  (Note that it is  possible for the
Lindblad operators to be different during the operation of each gate.)

Let us now use this expansion to analyze the effectiveness of error
correction in the presence of noise.  For simplicity we will examine
the three-qubit error correction scheme for dephasing noise.

As we see from figure 1 the 3-qubit encoding scheme can be effected by
a sequence of 5 Hamiltonians given in section~\ref{dephnoise}:
\begin{equation}
\Tenc(\Encod,0) = \e^{-i\H_5 t_5} \e^{-i\H_4 t_4} \e^{-i\H_3 t_3}
  \e^{-i\H_2 t_2} \e^{-i\H_1 t_1},
\end{equation}
while the decoding scheme is effected by applying the same gates in
the opposite order:
\begin{equation}
  \Tdec(\Decod,0) = \e^{-i\H_1 t_5} \e^{-i\H_2 t_4} \e^{-i\H_3 t_3}
  \e^{-i\H_4 t_2} \e^{-i\H_5 t_1}.
\end{equation}
In this case, $\Encod = \Decod = t_1 + t_2 + t_3 + t_4 + t_5$.  The
sum of these times is the time $\Encod+\Decod=\Delta$ defined above.  The
effects of the environment are summarized by three Lindblad operators
of the form $\sqrt{\kappa}\sz$, one for {\em each\/} qubit.  The
effective Hamiltonians then become
\begin{equation}
  \Heff = \H - {3 i \kappa \over 2} \one.
\end{equation}
Assume that the qubits evolve for a time $T$ between error correction
steps, defined as the transmission scenario in Sect.~\ref{transm}.
The procedure is then as follows: the initial qubit is encoded into
three qubits in a time $\Encod$, evolves for a time $T$ undisturbed,
and is then decoded in time $\Decod$.  Assume further that between the
encoding and decoding stages, the Hamiltonian $\H_0 = 0$ for the three
qubits (interaction picture).  The density operator is then
\begin{eqnarray}
\rhat(T+\Delta) = && \e^{-3\kappa(T+\Delta)} \biggl( \rhat(0) \nonumber\\
&& + \sum_{j=1}^3 \int_0^T \left[ \Tdec \L_j \Tenc \rhat(0)
  \Tenc\d \L_j\d \Tdec\d \right] dt \nonumber\\
&& + \sum_{j=1}^3 \int_0^{\Delta/2} \left[\Tdec\Tenc(\Delta/2,t)\L_j \Tenc(t,0)
  \rhat(0) \right. \nonumber\\
&& \left. \times \Tenc\d(t,0) \L_j\d \Tenc\d(\Delta/2,t) \Tdec\d
\right] dt \nonumber\\
&& + \sum_{j=1}^3 \int_0^{\Delta/2} \left[\Tdec(\Delta/2,t)\L_j\Tdec(t,0) \Tenc
  \rhat(0) \right. \nonumber\\
&& \left. \times \Tenc \Tdec\d(t,0) \L_j\d \Tdec\d(\Delta/2,t) \right] dt \nonumber\\
&& + \cdots \biggr).
\end{eqnarray}
The third and fourth terms of this expansion represent the possibility
of an error occurring {\it during} the encoding or decoding phase.
Such errors cannot necessarily be corrected, and represent a loss
additional to that from the higher order terms in the expansion.  The
first two terms represent the possibilities of no errors or a single
correctable error occurring.

The longer the time $T$ between error correcting steps, the larger the
higher-order terms, while the shorter the time $T$, the higher the
proportion $\delta = \Delta/T$ spent encoding and decoding.  Hence,
there should be an optimal time $T_{\rm opt}$ between error
corrections as a function of $\kappa$ and $\Delta$, which minimizes
the total error rate, consistent with the result of
(\ref{topt}).  For dephasing or isotropic noise, (\ref{topt}) will
hold exactly to lowest order.

\section{Numerical simulation}
\label{sec4}

Since the dimension of the Hilbert space of $n$ qubits is $2^n$, the
density operator $\hat\rho$ in the $n$-qubit case can be represented
by a complex Hermitian $2^n\times2^n$ matrix. This puts severe
constraints on the memory of the computer used to simulate these
systems. For $n=3$ and $n=5$, a direct numerical solution of the
master equation (\ref{eqmaster}) is feasible on a workstation.
Simulating the master equation for larger values of $n$ requires a
much larger computer (more memory in particular), because of the
exponential growth of the Hilbert space dimension.  The difficulty of
simulating such small $n$ systems on a {\it classical} computer
illustrates how much power would exist in a real quantum machine,
where the computation would actually run in the Hilbert space.

Our $n=3$ results (see below) were obtained by a straightforward
integration of the density matrix, using a fifth-order Runge-Kutta
algorithm for the numerical integration.  For our $n=5$ calculations,
we have used an alternative approach, the {\it quantum state
diffusion\/} method. This involves an average over sequential
evolutions of a $2^{n}$-dimensional quantum state, and therefore needs
less computer memory. A workstation can probably handle up to about a
dozen qubits if they are simulated this way.  The drawback is that,
because of the sequential runs required to construct good statistics,
such a simulation may require a lot of CPU time.  The $n=5$ case can
be handled by direct integration of the master equation and we have
checked the accuracy of our quantum state calculations using direct
integration of the master equation. Since the required computer memory
grows like $2^{2n}$ for a direct solution of the master equation, but
only like $2^n$ for a quantum trajectory simulation, the latter method
can be used for values of $n$ where the former would be impractical.

Before we present our numerical results, we give a short description of
quantum trajectories as numerical methods, with particular reference to
quantum state diffusion.

\subsection{Quantum trajectory simulations}

The storage problem due to large density matrices can be overcome by
unraveling the density operator evolution into {\it quantum
  trajectories\/} {\cite{Diosi1986, Gisin1992c, Carmichael1993b,
    Dalibard1992, Gardiner1992, Breslin1995}}.  Since quantum
trajectories represent the system as a state vector rather than a
density operator, they often have a numerical advantage over solving
the master equation directly, even though one has to average over many
quantum trajectories to recover the solution of the master equation. A
single quantum trajectory can  also give an excellent, albeit qualitative,
picture of a single experimental run.

We see from section \ref{secerrorqj} that we can justify the use of
the stochastic error models in sections 2 and 3, in spite of the
continuous, deterministic character of the master equation itself.
This type of treatment, in which the evolution of the density operator
is written as a sum over many different stochastic evolutions of
single wavefunctions, is called an {\it unraveling} of the master
equation, and a single realization of these evolutions is a {\it
  quantum trajectory}.  The unraveling of section 3.4 is often used in
simulating quantum optical systems, where it is known as the ``Quantum
Jumps'' or ``Monte Carlo Wavefunction'' approach
\cite{Carmichael1993b, Dalibard1992, Gardiner1992}.

The evolution of a single quantum jumps trajectory is given by the
(It\^o) stochastic differential equation
\begin{eqnarray}
\ket{d\psi} = && - i \hat H \,\ket\psi dt
  - {1\over2} \sum_j \left( \L\d_j \L_j
  - \expect{\L\d_j \L_j}_\psi \right) \ket\psi dt \nonumber\\
&& + \sum_j \left(
  {\L_j \over \sqrt\expect{\L\d_j \L_j}} - 1 \right) \ket\psi dN_j
\label{qjeqeq}
\end{eqnarray}
where the $dN_j$ are real stochastic differential variables which are
$0$ except at certain random times when they assume the value $1$.
These are independent, such that $dN_j dN_k = \delta_{jk} dN_j$, and
have a mean rate of jumps ${\rm M}(dN_j) = \expect{\L\d_j \L_j}_\psi
dt$.  Angular brackets denote the quantum expectation $\langle\hat
G\rangle_\psi = \langle \psi|\hat G|\psi\rangle$ of the operator $\hat
G$ in the state $|\psi\rangle$. The evolution between jumps is
continuous and differentiable.  The density operator is given by the
mean over the projectors onto the quantum states of the ensemble.  If
the pure states of the ensemble satisfy the equation (\ref{qjeqeq}),
then the density operator given by
\begin{equation}
  \hat\rho(t)= {\rm M}\ket{\psi(t)} \bra{\psi(t)},
\label{densopeq}
\end{equation}
satisfies the master equation~(\ref{eqmaster}).
{From} this it is clear that the expectation value of an operator $\hat
O$ is given by
\begin{equation}
  \Tr\{\hat O\hat\rho \} = {\rm M} \bra\psi\hat O \ket\psi \;.
\end{equation}

Quantum jumps is a useful conceptual picture, but it is not the only
unraveling of the master equation.  It is convenient that
we can use whatever unraveling we choose based solely on calculational
convenience, as they are all equivalent to the master equation.  Among
the most important is the quantum state diffusion (QSD) equation of
Gisin and Percival {\cite{Gisin1992c}}.  We have applied both jump and
QSD equations to the problems considered in this paper. It turns out
that to obtain good statistics, a significantly smaller number of
trajectories need be summed when the QSD equation was used.  We thus
limit further discussion to the QSD equation, a nonlinear stochastic
differential equation for a normalized state vector $|\psi\rangle$:
\begin{eqnarray}
|d\psi\rangle &=& -{i\over\hbar} \hat H \,|\psi\rangle dt 
  + \sum_j\left(\langle \hat L_j^\dagger\rangle_\psi \hat L_j
  - {1\over2} \hat L_j^\dagger \hat L_j 
  - {1\over2} \langle \hat L_j^\dagger\rangle_\psi
  \langle\hat L_j\rangle_\psi\right) |\psi\rangle dt \nonumber \\
&&+\sum_j \left(\hat L_j -\langle \hat L_j\rangle_\psi\right) 
          \,|\psi\rangle d\xi_j \;.
\label{eqqsd}
\end{eqnarray}
The first sum in this equation represents the deterministic drift of
the state vector due to the environment, and the second sum the random
fluctuations. 
The $d\xi_j$ are
independent complex differential Gaussian random variables satisfying
the conditions
\begin{equation}
  {\rm M} d\xi_j = {\rm M} d\xi_i d\xi_j = 0\;,\;\;\; {\rm M} d\xi_i^*
  d\xi_j = \delta_{ij}dt \;,
\end{equation}
where {\rm M} denotes the ensemble mean.  A QSD trajectory is
continuous, but not differentiable.  If the pure states of the
ensemble satisfy the QSD equation~(\ref{eqqsd}), then the density
operator given by (\ref{densopeq}) again satisfies the master
equation~(\ref{eqmaster}).

To simulate the QSD equation, we use a publicly available C++ software
library written by two of the authors \cite{cpplib}. The software uses
object-oriented programming concepts to allow great flexibility in
defining operators and states in Hilbert spaces with arbitrary numbers
of degree of freedom. As an illustration, we show how the list of
Hamiltonians that defines the network effecting the encoding is
implemented:
\begin{verbatim}
const int nOfGates=10;
Operator H[nOfGates] = { A1+A2+A3, B023, C023, A4, B04, A0,
                         B03+B14, B01+B24, A0+A4, B03 };
\end{verbatim}
The operators in the list are implemented as follows (again we give as an
example the operators {\tt A2} and {\tt B04} only):
\begin{verbatim}
  IdentityOperator id;
  SigmaX sx2(2);
  SigmaZ sz0(0);
  SigmaZ sz2(2);
  SigmaZ sz4(4);
  Operator pr4 = 0.5*(id+sz4);
  Operator pr0 = 0.5*(id+sz0);
  Operator A2 = (M_PI/2)*(sqrt(0.5)*(sx2-sz2) + id);
  Operator B04 = (M_PI)*pr0*pr4;
\end{verbatim}

\subsection{Numerical Results}

The results obtained with the QSD method have been checked against
those obtained by a direct integration of the master equation. The
disagreement, of the order of 1-2\% is purely statistical and is due to
the finite number of trajectories used to build the average
(around 200 trajectories for the three bit code
and 200--400 for the five bit code).

The simulations confirm the analytical results discussed in the
previous section, both for the three-bit and the five-bit code.  One
measure of the efficiency of a quantum error correcting code is the
mismatch between the decohered, corrected ensemble and the initial
state, as defined in (\ref{mismatchdef}).  This mismatch indicates
how faithfully the initial state has been preserved in the face of noise.

\begin{figure}
\vspace{4mm}
\centerline{\psfig{width=6in,file=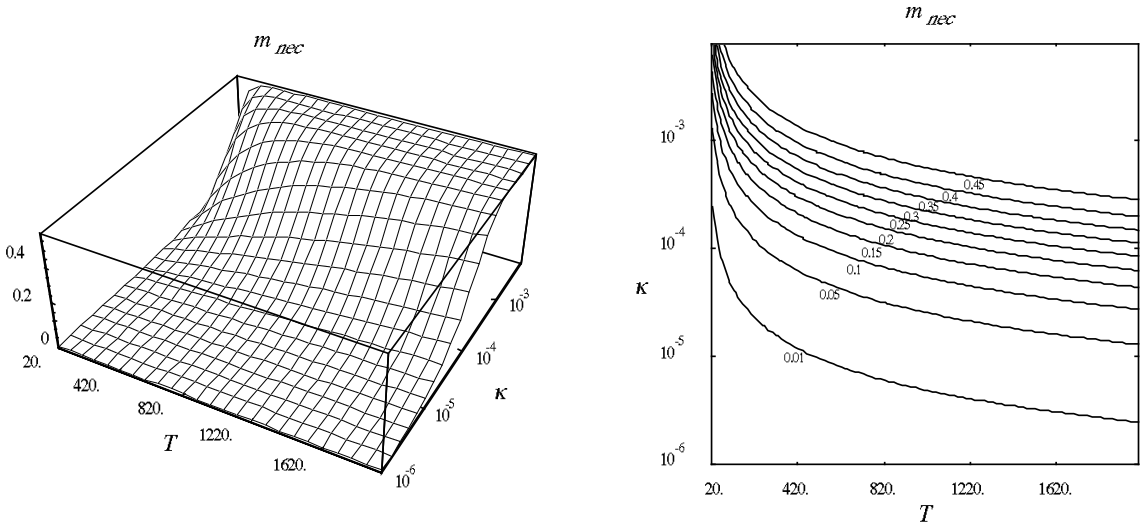}}
\vspace{2mm}
\caption[fo2]{
  \small Mismatch between the initial and final states of a qubit
  in contact with an isotropic noise reservoir. The time
  scale starts at $t=20$ to match the encoding and
  decoding times in comparing this figure with the numerical results
  for the case of a qubit with error correction. This follows
  the storage scenario of Sect~\ref{storagesec}.  }
\label{figm1}
\end{figure}

The mismatch $m_{nec}$ for a single qubit undergoing decoherence is defined
by (\ref{misphas}) for phase noise and (\ref{misiso}) for isotropic
noise.  Fig.~\ref{figm1} shows the isotropic noise mismatch of
a single qubit.  This is the benchmark to evaluate the
efficiency of the five bit error correction code.

\begin{figure}
\vspace{4mm}
\centerline{\psfig{width=6in,file=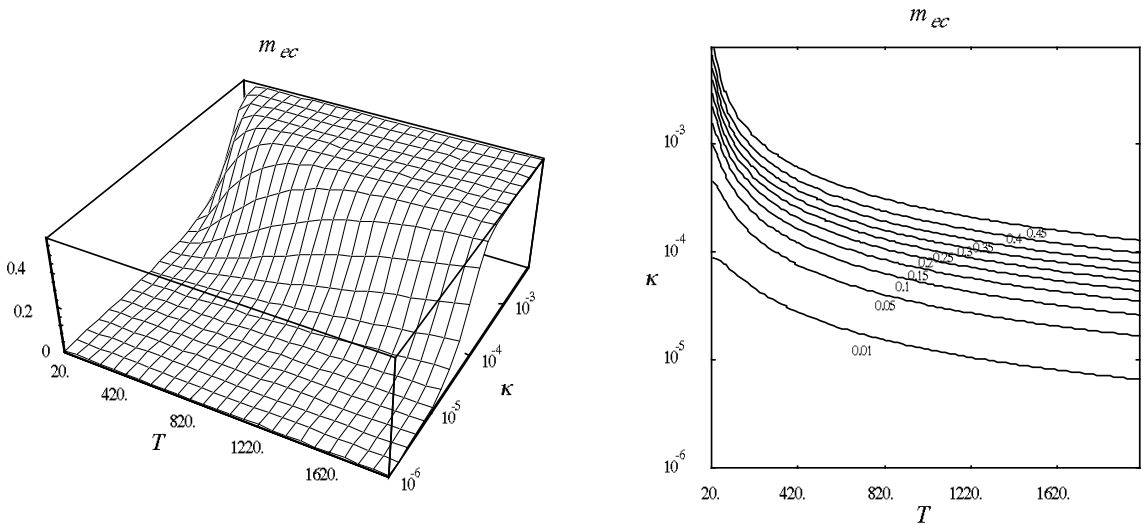}}
\vspace{2mm}
\caption[fo3]{
  \small Mismatch between the initial and final states of a
  qubit with a five bit error correction code.  The time scale starts
  at $t=20$ units to include the encoding and decoding
  times ($\Delta=10+10=20$).  Each step (one or several gates)
  is effected in unit time, as indicated in Fig.~\ref{5bitfig}.}
\label{figm5}
\end{figure}

A similar figure can be obtained by plotting the mismatch $m_{ec}$ of
a qubit that has been encoded and later decoded.
Instead of looking at the mismatch of a single qubit in contact with
an isotropic noise reservoir for a time $T$, one encodes the qubit into
five qubits
($10$ units of time), allows the five qubits to interact with the reservoir
for $T-20$ units of time, and finally decodes and corrects the qubit
($10$ units of time). The resulting  reduced density operator is used to
compute $m_{ec}$ via (\ref{mismatchdef}). This mismatch, obtained by
numerical simulation, is illustrated in Fig.~\ref{figm5}.

\begin{figure}
\vspace{4mm}
\centerline{\psfig{width=3in,file=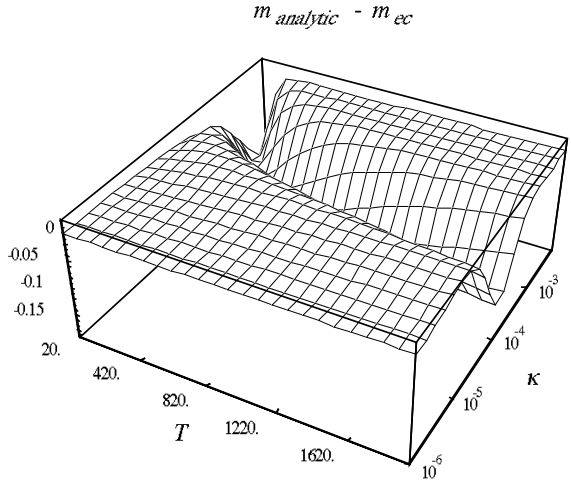}}
\vspace{2mm}
\caption[fo4]{
  \small Difference between $m_{ec}$ and $m_{analytic}$, where $m_{ec}$
  is the mismatch occuring in the numerical simulation and $m_{analytic}$
  is the estimated mismatch from section 3.  In both cases a single
  qubit is encoded into five and allowed to evolve in contact with a
  noisy environment, with noise occurring also during the
  encoding and the decoding phases.  Time on the axis represents
  encoding and decoding time (i.e., $\Delta=10+10=20$ units of time)
  plus the time the encoded qubit is left interacting with
  the environment. }
\label{disagreement5bit}
\end{figure}

Alternatively, we can use the results of Sect.~\ref{analytic} and
estimate the mismatch by
\begin{equation}
  m_{analytic}=\frac{1}{2} (1 - s_{sc}(5)),
\label{manalytic}
\end{equation}
where $s_{sc}(n)$ is defined in (\ref{pstore}).  The agreement
between this simple analytical model and the numerical simulations is
very good, as illustrated in Fig.~\ref{disagreement5bit}. The maximum
discrepancy is of the order of 10 percent for our range of parameters.

Consistent with the analysis developed in Sect.~\ref{secres}, one
can identify in these numerical simulations a region of the $\kappa$-$T$
plane for which error correction is likely to help, despite noise
occurring in the encoding and decoding stages. This can be clearly
seen by looking (in analogy with Eq.~\ref{firstratio}) at the positive
values of $\log(m_{nec}/m_{ec})$.

\begin{figure}
\vspace{4mm}
\centerline{\psfig{width=6in,file=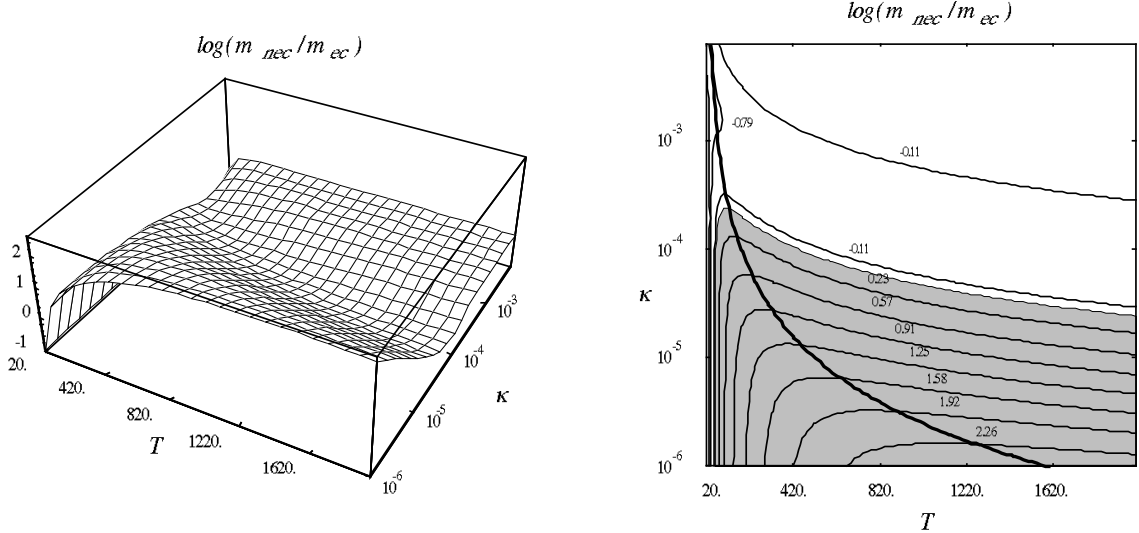}}
\vspace{2mm}
\caption[fo5]{
  \small Surface and contour plots of $\log(m_{nec}/m_{ec})$, where
  $m_{ec}$ is generated numerically. The shaded area indicates values of
  $(\kappa,T)$ for which error correction is useful. This figure
  should be compared to Fig.~\ref{logplotcalc}. The thick line
  represents, for each given $\kappa$, the optimum time between the
  \Encod\ and \Decod\ stages (cf.  Sect.~\ref{secres},
  Eq.~\ref{topt}). }
\label{logplot}
\end{figure}

\begin{figure}
\vspace{4mm}
\centerline{\psfig{width=6in,file=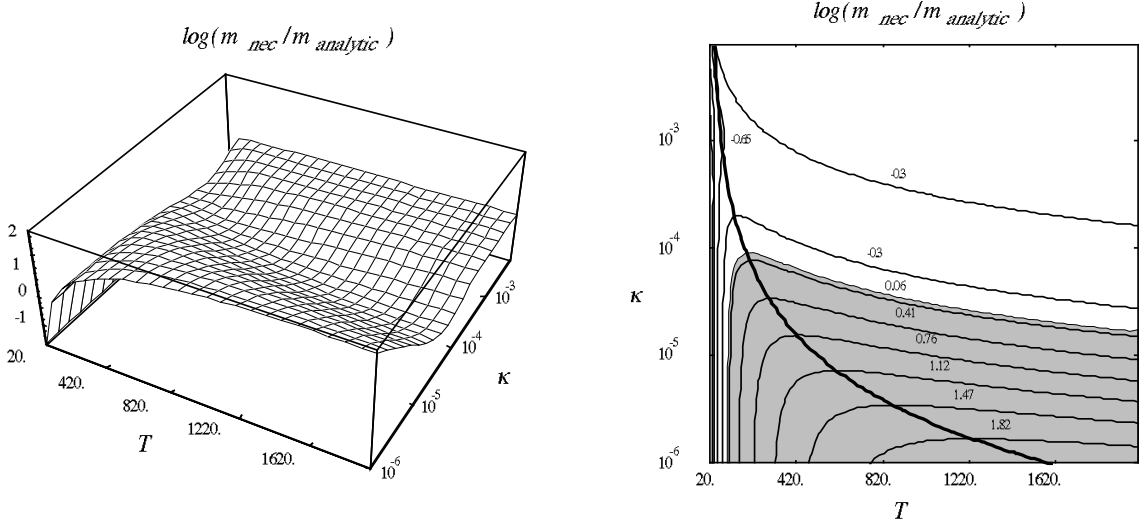}}
\vspace{2mm}
\caption[fo6]{
  \small \small Same as Fig.~\ref{logplot}, but here we plot
  $\log(m_{nec}/m_{analytic})$ instead of $\log(m_{nec}/m_{ec})$.  The
  discrepancy between the numerical calculations and the analytical
  arguments is illustrated in Fig.~\ref{disagreement5bit}.}
\label{logplotcalc}
\end{figure}

Another possible measure of the benefit of
error correction is the difference $m_{ec}-m_{nec}$. Positive values
of the difference indicate that error correction is worthwhile.
However, this measure is of little use when both $m_{ec}$ and $m_{nec}$
go to zero, as their difference also vanishes. The $\log$ of the ratio
does not have this drawback, and is therefore preferable as an
indicator of where error correction is beneficial.
The  region of positive values of the $\log$ is represented by
the shaded area in Fig.~\ref{logplot}. One notices, as expected, that
for small enough $\kappa$ and large $T$, error correction is desirable
(since $\log(m_{nec}/m_{ec})>0 \iff m_{ec}<m_{nec}$). For
comparison, Fig.~\ref{logplotcalc} shows the same quantity where the
analytical expression $m_{analytic}$ of (\ref{manalytic}) has been
used instead of $m_{ec}$.

An exactly similar set of calculations can be done for three bit codes
in the case of delocalizing noise, and similar behavior was observed.
In both the three bit and five bit cases, there is a section of the
$\kappa$-$T$ plane where error correction remained beneficial even in
the presence of noise during encoding and decoding; and for low values
of the environmental interaction strength $\kappa$, there was an optimal
time between error correction steps.  This is consistent with the result
obtained by Chuang and Yamamoto \cite{chuang}.

\section{Conclusions}

{From} both the analytical arguments and the numerical simulations, we
see that error correction can prove worthwhile even in the
presence of noise during encoding and decoding.  For a
given strength of the environmental coupling, there is an optimal rate
at which error correction should be performed, and for a given time of
storage there is an optimal number of error correction steps.

The simulations presented in this paper are among the first to treat
both the execution of gates and the influence of the environment
realistically, in the sense that the operation of gates takes a finite
amount of time during which noise continues to act on the system.
Moreover, models of the noise were used which correspond to common
environmental effects in atomic and optical physics.

While the theory of error correction has moved rapidly, it is unlikely
that circuits involving many qubits will be experimentally realized soon.
Systems of a few qubits thus remain of great interest.
Three-bit and five-bit error correction are among
the first circuits that might be experimentally implemented,
and hence our results should be of relevance to
near-future experiments in this field.

We also have seen that quantum trajectories provide a practical
technique for simulating systems with multiple qubits.  This may prove
particularly useful in treating systems with many qubits, where
solving the full master equation is impractical due to the large size
of the Hilbert space.

These simulations could be improved and extended in many ways.
The Hamiltonians used to represent the gates were chosen for
convenience rather than reflecting any particular physical system.  It
would be useful to get closer to the actual physics of proposed
quantum computers, such as the linear ion trap of Cirac and Zoller
\cite{cirac}.  In the same way, the coupling to the environment might
differ for different gates.  It would be straightforward to include
these effects.

There are other interesting problems in quantum computation which
might be studied by techniques like those of this paper.  The recently
proposed fault-tolerant error correction schemes are far more
complicated than the ones treated in this paper.  They would be beyond
the reach of direct numerical simulation of the master equation with
present computers, but may well prove amenable to a quantum
trajectory approach.

Quantum computation still faces many hurdles before becoming reality.
But it is far too early to say that the ingenuity of those working in
the field is not sufficient to overcome them.

{\bf Acknowledgments}

The authors acknowledge A. Ekert, N. Gisin, R. Laflamme, I.C. Percival
and M. Plenio for useful conversations.  We are grateful to
Hewlett-Packard for use of computational resources. A.B. acknowledges
the financial support of the Berrow's fund at Lincoln College (Oxford)
and of the Swiss National Science foundation.  T.A.B. and R.S.
acknowledge financial support from the UK EPSRC.  This research was
partially supported by the TMR Network Programme of the European
Commission on the Physics of Quantum Information.

\end{document}